\documentclass[%
 aip,
rsi,%
 amsmath,amssymb,
 reprint,%
]{revtex4-2}

\usepackage[dvipsnames,table]{xcolor}
\usepackage{xcolor}
\usepackage{graphicx}% Include figure files
\usepackage{dcolumn}% Align table columns on decimal point
\usepackage{bm}% bold math
\usepackage{nicefrac}
\newcommand{\n}[1]{\mathbf{#1}}

\begin{document}

\preprint{AIP/123-QED}

\title[Rodr\'{i}guez et al.]{Weakly Quasisymmetric Near-Axis Solutions to all Orders}% Force line breaks with \\

\author{E. Rodr\'{i}guez}
 \altaffiliation[Email: ]{eduardor@princeton.edu}%Lines break automatically or can be forced with \\
 \affiliation{ 
Department of Astrophysical Sciences, Princeton University, Princeton, NJ, 08543%\\This line break forced with \textbackslash\textbackslash
}
\affiliation{%
Princeton Plasma Physics Laboratory, Princeton, NJ, 08540%\\This line break forced% with \\
}%

\author{W. Sengupta}
%  \altaffiliation[Email: ]{amitava@princeton.edu}
 \affiliation{ 
Department of Astrophysical Sciences, Princeton University, Princeton, NJ, 08543%\\This line break forced with \textbackslash\textbackslash
}
\affiliation{%
Princeton Plasma Physics Laboratory, Princeton, NJ, 08540%\\This line break forced% with \\
}%

\author{A. Bhattacharjee}
 \altaffiliation[Email: ]{amitava@princeton.edu}
 \affiliation{ 
Department of Astrophysical Sciences, Princeton University, Princeton, NJ, 08543%\\This line break forced with \textbackslash\textbackslash
}
\affiliation{%
Princeton Plasma Physics Laboratory, Princeton, NJ, 08540%\\This line break forced% with \\
}%

\date{\today}% It is always \today, today,

\begin{abstract}
We show that the equations satisfied by weakly quasisymmetric magnetic fields can be solved to arbitrarily high order in powers of the distance from the magnetic axis. {  This demonstration does not consider force balance.} The existence of solutions requires an appropriate choice of parameters, most notably the toroidal current or rotational transform profiles. We do not prove that the expansion converges (it is likely {  divergent} but asymptotic), and thus the demonstration here should not be taken as definitive proof of the existence of global solutions. Instead, we provide a systematic construction of solutions to arbitrarily high order. 
\end{abstract}

\maketitle

\section{Introduction:}\label{sec:intro}
The notion of \textit{quasisymmetry} (QS) as a basis for magnetic confinement design has been promising for a few decades now. Initially, QS emerged as a requirement on the magnetic field so that the neoclassical behavior of the stellarator is equivalent to that of an axisymmetric tokamak.\cite{boozer1983,nuhren1988} This allowed for the design of 3D stellarators which do not suffer excessively from unfavorable confinement due to $1/\nu$ transport.\cite{Helander2014} \par
However, it was soon realized that this QS property had some limitations. In a seminal paper, Garren and Boozer\cite{garrenboozer1991b} suggested that construction of configurations that bear QS and satisfy the magnetohydrostatic (MHS) equations everywhere does not appear possible. They did so by showing that the governing equations lead to an overdetermined system when expanded in powers of the distance from the magnetic axis.\par
It was not clear whether the overdetermination conundrum originated in the idea of QS itself or whether it was a consequence of the insistence on magnetohydrostatics. To uncover the root of the problem, renewed emphasis was recently placed on reformulating the concept of QS in a way that made no assumptions regarding the underlying equilibrium.\cite{burby2020,rodriguez2020}  A field is considered \textit{weakly quasisymmetric}\cite{rodriguez2020} iff it guarantees the existence of an approximately conserved momentum for the dynamics of guiding-center motion. Such a separation of the concept of QS from considerations of magnetohydrodynamic (MHD) equilibria allows one to avoid the overdetermination problem in the presence of macroscopic forces that are more general than isotropic pressure gradients\cite{rodriguez2020i,rodriguez2020ii,constantin2021,rodrigGBC}. \par    
Although relaxation of equilibrium considerations avoids the Garren-Boozer overdetermination problem, it has not been shown that the ordered set of equations for QS (without invoking any assumptions on the nature of the underlying equilibrium) may actually be solvable. In this paper, we prove that one can do so for a weakly quasisymmetric field and to arbitrarily high order. This is a first step in demonstrating that no inherent limitation seems to be associated with realizing weak QS. The constructive proof presented here will also serve in the future as a test case for analysis of the convergence (or asymptotic) properties of the near-axis expansion of the solution to arbitrarily high order. \par
The paper is organized as follows. In Section II, we present the governing equations, the formalism for the near-axis expansion, and a succinct account of the expanded equations and how to solve them. This draws heavily from [\onlinecite{rodriguez2020i}], to which we refer for additional details. The systematic solutions to the equations are presented in Appendices. Section III focuses on a recurring ordinary differential equation: the generalized $\sigma$-equation. The equation is analyzed and solved, and the existence of solutions is shown to be related to the determination of current and rotational transform profiles. Next, we discuss in Section IV how to formulate this problem appropriately, emphasizing the functions meant to be inputs and those that are outputs, { and present a numerical example}. Finally, we conclude in Section V with a summary.

\section{Near-axis expansion: principle and governing equations}
Our primary purpose here is to try and construct magnetic field solutions that are quasisymmetric in the weak sense\cite{rodriguez2020,burby2020}. Doing so implies solving the appropriate set of governing equations, often referred to as \textit{magnetic} equations.\cite{rodriguez2020i,rodrigGBC} We shall briefly introduce these multiple, coupled partial differential equations (PDEs) in this Section. \par
Generally, the complexity of the problem is forbidding. Hence it has been common practice\cite{garrenboozer1991a,garrenboozer1991b,landreman2018a,landreman2019,rodriguez2020} to consider the problem perturbatively. The idea of near-axis expansions consists of taking the set of governing equations and associated functions and expanding them in powers of the distance from the magnetic axis.\cite{Mercier1964,Solovev1970,garrenboozer1991a} Treating that distance as an ordering parameter reduces the global set of PDEs into a hierarchy of simpler equations. In the case that concerns us here, the equations become algebraic and ordinary differential equations (ODEs). Complete details on proceeding with such expansions can be found in the cited literature, especially [\onlinecite{rodriguez2020i}]. After constructing the magnetic equations, we present the fundamentals of the expansion procedure and the structure of the expansion. 
\par
\subsection{Magnetic equations}
There are two principal constraints that we need to impose on the vector field $\mathbf{B}$: first, its solenoidal nature, and second, QS. This part of the problem we refer to as the \textit{magnetic} part\cite{rodriguez2020i,rodrigGBC}, distinguishing it from the \textit{equilibrium} part. \par
We choose a coordinate set to express these constraints that will enable us to carry out the near-axis expansion later straightforwardly. It is convenient to choose \textit{generalized Boozer coordinates} (GBC)\cite{rodrigGBC}. GBCs are an extension of traditional Boozer coordinates. These new coordinates do not require $\mathbf{j}\cdot\nabla\psi=0$, but do the weaker condition $\oint(\mathbf{j}\cdot\nabla\psi)\mathrm{d}l/B=0$ along field lines, property that is guaranteed by weak QS. The three GBCs are written $\{\psi,\theta,\phi\}$, which denote toroidal flux, poloidal angle, and toroidal angle, respectively. We then write the governing equations in their straight-field-line form. To implement the particular form of the coordinates, in addition, the Jacobian needs to be prescribed. \par
The straight-field-line character can be incorporated by writing the magnetic field in its contravariant form\cite{rodrigGBC},
\begin{align}
    \n{B}=\nabla\psi\times\nabla\chi+\bar{\iota}\nabla\phi\times\nabla\psi, \label{eqn:Bcontr}
\end{align}
where $\bar{\iota}=\iota-N$, $\iota$ is the rotational transform and $N$ is an integer. Here the helical coordinate $\chi=\theta-N\phi$ is introduced for convenience. The Jacobian is then prescribed as $\mathcal{J}=(\nabla\psi\times\nabla\theta\cdot\nabla\phi)^{-1}=B_\alpha(\psi)/B^2$, where $B_\alpha$ is a flux function (which, in standard Boozer coordinates, is $G+\iota I$, with $G$ and $I$ flux functions representing Boozer currents) and $B$ is the magnitude of the magnetic field. \par
The presence of $B$ in the Jacobian in an explicit form is very convenient. The magnetic field is weakly quasisymmetric iff the magnetic field magnitude is of the form $B=B(\psi,\chi)$.\cite{rodrigGBC,rodriguez2020} Thus, QS can be imposed by requiring the Jacobian to have the appropriate dependence only on the angular coordinate $\chi$. In this context, the integer $N$ determines the class of symmetry we are dealing with: $N=0$ for quasi-axisymmetry (QA) and otherwise for quasi-helical symmetry (QH). In this formulation, $B$ naturally appears as an input to the problem. Cast in this form, the formulation generalizes the classical form\cite{boozer1983,rodrigGBC} and is less constraining than its strong form\cite{tessarotto1996,burby2019,rodriguez2020}. \par
Using the same coordinates, the magnetic field may also be written in its covariant form,
\begin{equation}
    \mathbf{B}=B_\theta\nabla\chi+(B_\alpha(\psi)-\bar{\iota}B_\theta)\nabla\phi+B_\psi\nabla\psi. \label{eqn:Bcov}
\end{equation}
\par
The considerations above complete the governing equations for the weakly quasisymmetric magnetic field problem. Equations (\ref{eqn:Bcontr}) and (\ref{eqn:Bcov}) are consistent with the solenoidal constraint on the magnetic field and the existence of magnetic surfaces. Requiring the Jacobian to be of the GBC form, with $B$ given as an input, ensures QS. \par
Given these equations, there may be multiple strategies to solve them. %Interpreting the equations as constraints for the covariant components and the GBC coordinates is not convenient--amongst other things, the $\chi$ dependence of $B$ would difficult the form of the governing PDEs).% 
Treating the magnetic coordinates $\{\psi,\theta,\phi\}$ as independent is a convenient way to formulate a solution.
This approach is known as the \textit{inverse} formulation. The position vector $\mathbf{x}$ is expressed in the Frenet-Serret basis associated with the magnetic axis \cite{garrenboozer1991a}:
\begin{equation}
    \n{x}-\n{r}_0(l)=X\hat{\kappa}(l)+Y\hat{\tau}(l)+Z\hat{b}(l). \label{eqn:inverseMap}
\end{equation} 
Here $\{X,~Y,~Z\}$ are functions of the generalized Boozer coordinates, $\n{r}_0$ represents the position of the magnetic axis, and $l$ {  (a function of $\phi$)} parametrizes the distance along it. The Frenet-Serret basis is given by $\hat{b}$, $\hat{\kappa}$ and $\hat{\tau}$, along with the directions of the tangent, the normal and binormal vectors to the magnetic axis, respectively. The description of the magnetic axis is encoded in $\mathbf{r}_0$, which can be uniquely described by its curvature, $\kappa$, and torsion, $\tau$. \par
It is convenient to employ the \textit{dual relations}\cite{garrenboozer1991a} to write, e.g., $\nabla\psi\times\nabla\chi=\mathcal{J}^{-1}\partial\n{x}/\partial\phi$ and $\nabla\phi=\mathcal{J}^{-1}\partial\n{x}/\partial\psi\times\partial\n{x}/\partial\chi$. With the role of the GBCs properly established, we may rewrite the two governing magnetic equations in the form presented in [\onlinecite{rodriguez2020}]. First, we write the \textit{Jacobian} equation, denoted by $J$,\cite{rodriguez2020i}
\begin{equation}
    \frac{B_\alpha(\psi)^2}{B(\psi,\chi)^2}=\left|\frac{\partial\n{x}}{\partial \phi}+\Bar{\iota}\frac{\partial\n{x}}{\partial \chi}\right|^2. \label{eqn:Jgen}
\end{equation} 
The other equation is the \textit{co(ntra)variant} equation. This equates the covariant and contravariant forms Eqs.~(\ref{eqn:Bcov}) and (\ref{eqn:Bcontr}). Using the dual relations, we obtain
\begin{align}
    (B_\alpha-&\Bar{\iota}B_\theta)\frac{\partial\n{x}}{\partial\psi}\times\frac{\partial\n{x}}{\partial\chi}+B_\theta\frac{\partial\n{x}}{\partial\phi}\times\frac{\partial\n{x}}{\partial\psi}+B_\psi\frac{\partial\n{x}}{\partial\chi}\times\frac{\partial\n{x}}{\partial\phi}=\frac{\partial\n{x}}{\partial\phi}+\nonumber\\
    &+\Bar{\iota}(\psi)\frac{\partial\n{x}}{\partial\chi}. \label{eq:co(ntra)variant}
\end{align}
This equation has three scalar projections along the Frenet-Serret basis. These components are denoted by $C_b$, $C_\kappa$ and $C_\tau$. The two latter components generally appear together and are referred to as $C_\perp$. \par
\par
\subsection{Near-axis expansion}
We are now set up to introduce the near-axis expansion procedure. The approach consists on expanding all $\psi$ dependent functions describing the magnetic field in powers of the distance to the magnetic axis. For this purpose, we define a pseudo-radial coordinate $\epsilon=\sqrt{(\kappa_\mathrm{max})^2\psi/B_\mathrm{min}}$, where $\psi=0$ on axis, $\kappa_\mathrm{max}$ is the maximum curvature on axis and $B_\mathrm{min}$ the minimum magnetic strength on axis. (Note that the normalisation of $\epsilon$ has variations in the literature but these are not germane to the expansion process. For simplicity, we shall take $\epsilon=\sqrt{\psi}$.)   {  Considering that the functions expanded are free of singularities in the neighbourhood of the axis,} spatial functions may then be expanded in a Taylor series. {  To do so appropriately using the pseudo-radial coordinate $\epsilon$}, integer powers of $\epsilon$ must be carefully combined with a Fourier expansion in the angle $\chi$, so that any function $f$ may be written in the form,\cite{landreman2018a,garrenboozer1991a,kuo1987}
\begin{equation}
    f=\sum_{n=0}^\infty\epsilon^n{\sum_{m=0|1}^{n}}\left[f_{nm}^c(\phi)\cos m\chi+f_{nm}^s(\phi)\sin m\chi\right].
\end{equation} 
The latter sum runs over even or odd indices depending on $n$ {  (the summand for each $n$ being represented by $f_n$)}. For our problem, an expansion of this form will apply to the covariant magnetic functions $B_\theta$ and $B_\psi$, as well as the spatial functions $X$, $Y$ and $Z$.  \par
The Fourier expansion presented is especially suited to imposing QS, as we may write 
\begin{equation}
    \frac{1}{B^2}=B_0+\sum_{n=1}^\infty\epsilon^n \sum_{m=0|1}^n\left(B_{nm}^c\cos m\chi +B_{nm}^s\sin m\chi\right),
\end{equation}
for the magnetic field strength, naturally including its symmetry requirement by having constant expansion coefficients. \par
Finally, there are two main flux functions involved in the problem, namely, $B_\alpha$ and $\bar{\iota}$. For these functions, the expansion reduces to,
\begin{equation}
    \iota(\psi)=\sum_{n=0}^\infty\epsilon^{2n}\iota_n.
\end{equation}
These series expansions are all we need to apply near-axis expansion to the magnetic equations (\ref{eqn:Jgen}) and (\ref{eq:co(ntra)variant}). One simply substitutes the expanded form of the various functions into the equations and collects terms sharing powers of $\epsilon$. In this method, information on the magnetic axis is contained in $\mathbf{r}_0$, which is a $\epsilon^0$ piece. Although this procedure, as sketched, is straightforward, the practical implementation involves lengthy algebra. We shall not be concerned with these details here, which have been dealt with carefully elsewhere\cite{rodriguez2020i}. Instead, we will focus on the general structure of the equations and the existence and solution of the resulting equations. The Appendices include systematic approaches to solve the expanded equations to arbitrarily high order. For notational convenience, hereafter, we label the expanded equations using a superscript on the equation label indicating the order of the expansion. For example, the second-order equations associated with the Jacobian equation are referred to as $J^2$ \par

\subsection{General structure of solution}
\begin{table}[]
    \centering \hspace*{-.5cm}
    \begin{tabular}{|c||c|c|c|}
        \hline
    Eqn. & Order & Solve for\dots & Nb. eqns \\ \hline\hline
    $J^n$ & $n$ & $X_n$ & $n+1$ \\\hline
    $C_b^n$ & $n$ & $Y_{n+1}$ & $n+1$ \\\hline 
    $C_\perp^n$ & $n=2k$ & $B_{\psi n-1}$ & $n$ \\
     &  & $Z_{n+1}$ & $n+2$ \\
     &  & $B_{\theta n+1,n+1}$ & $2*$ \\ \hline
    $C_\perp^n$ & $n=2k+1$ & $Y_n$ & $1"$ \\
     &   & $B_{\psi n-1}$ & $n-1$ \\
     &   & $Z_{n+1}$ & $n+2$ \\
     &   & $B_{\theta n+1,n+1}$ & $2*$ \\\hline
    \end{tabular}
    \caption{\textbf{Solution structure of magnetic equations.} Each column shows: the label of the expanded equation, the order of expansion, what the equations are solved for and the number of constraints (or independent equations) they amount to. The asterisk indicates that due to a trivial solution, two other constraints are also satisfied (but we drop them so that the total counting can be done correctly). The " indicates that a differential equation needs to be solved subject to periodic boundary conditions, unlike the rest of equations which simply require algebraic manipulations. This table is reproduced from [E.  Rodriıguez, A.  Bhattacharjee, Phys. Plasmas 28, 012508 (2021)], Table~I, with the permission of AIP Publishing.}
    \label{tab:countingQSJCaxis}
\end{table}
Table \ref{tab:countingQSJCaxis} summarizes how the expanded magnetic equations are to be solved for at every order (see also Table I in [\onlinecite{rodriguez2020i}]). This follows the analysis in [\onlinecite{rodriguez2020i}], and we will provide a constructive argument in the following paragraphs and Appendices. The key observation is that the majority of these equations are linear and algebraic. Thus their solution is generally straightforward, and holds to all orders. This is true with the exception of a recurrent ordinary differential equation that appears at every other order. This needs more careful consideration and will be the focus of the next Section. \par
For the remainder of this Section, we will consider the solution of the algebraic equations, which represent most of the constraints. While the near-axis expansion leaves us with a set of equations for which solutions may always be found, two essential aspects must be considered. First, we need to prescribe the input and output functions appropriately. Second, the solution should be shown to be uniquely obtainable from the information provided \par
The functions that can be solved for from algebraic equations can be read off from Table~\ref{tab:countingQSJCaxis}. Although our choice might seem arbitrary at first glance, it is possible to give a constructive argument for it. The formulation in this form gives a well-posed problem, as we shall see more explicitly in the Appendices and the description of the solutions to follow. To complement Table \ref{tab:countingQSJCaxis}, we shall specify the input functions. Let us start with the shape of the axis through $\kappa$ and $\tau$, which are prescribed as functions of $\phi$. The profiles of $\bar{\iota}$ (or the averaged toroidal current) and $B_\alpha$ also need to be specified (although considerations of the solvability of the recurring differential equation will be shown to limit the freedom of choice). Finally, the Fourier coefficients of the magnetic field magnitude need to be given, as must $\tilde{B}_\theta=B_\theta-\oint B_\theta\mathrm{d}\phi$. This latter $\tilde{B}_\theta$ represents the surface variations of the toroidal current and must abide by the relation $B_{\theta n n}=0$. The analysis of the expanded equations also leaves $Y_{n0}$ and $B_{\psi n0}$ unconstrained, which do affect the solution. Most notably, in MHS, $B_{\psi 0}'$ is related to the pressure gradient and with it to the Shafranov shift.  \par
Let us now spell out the solution structure of the expanded algebraic equations. We do so by looking at each of the solved functions at a time. 
\paragraph{$X_n$}: the Jacobian equation (\ref{eqn:Jgen}) is used to algebraically find a closed-form expression for $X_n$. The unknown $X_n$ may always be read off the $\epsilon^n$ component of the Jacobian equation $\mathcal{J}^n$ as shown in Appendix A. The equation is guaranteed to be invertible and free of singular behavior.
\paragraph{$Y_n$}: from the parallel component of the co(ntra)variant equation, $C_b^{n-1}$, we obtain a closed-form for $n$ of the $n+1$ components of the function $Y_n$. The structure of the solutions is significantly more involved than the $X_n$ construction, as presented in detail in Appendix B. However, the solution is guaranteed to exist and be well behaved. The construction leaves two important observations. First, that at even orders $Y_{n0}$ is left as a free function. And second, that $Y_{n1}^C$ is left to satisfy a differential equation at odd orders (indicated in Table~\ref{tab:countingQSJCaxis} for $C_\perp^n$ at odd orders). This differential equation is obtained from the $C_\perp$ equation, discussed in the next Section. To guarantee that solutions can be found systematically at all orders, we must show that the ODE is solvable.
\paragraph{$Z_n$}: the perpendicular components of the co(ntra)variant equation, $C_\perp$, can be used to systematically and algebraically obtain $Z_n$ in terms of known functions. Moreover, the algebraic equation does not contain any singular behavior (see Appendix C), and thus a solution to $Z$ will exist to all orders.
\paragraph{$B_{\psi n}$}: the other components of $C_\perp^n$ are used to find $B_\psi$. We provide brief outlines in Appendix D, which replicates the procedure in [\onlinecite{rodriguez2020i}]. Once again, $B_\psi$ can be constructed algebraically in terms of $B_\theta$ and other known functions. The expression for $B_\psi$ is well-behaved and unique (up to the undetermined $B_{\psi n0}$) to arbitrarily high order. The freedom in $B_{\psi n0}$ includes the relabelling of the origin of the coordinate $\phi$.\cite{landreman2019} 

From the above (and the explicit constructions presented in the Appendices), it should be clear that most of the work involves solving algebraic equations. These equations can be solved for $X$, $Y$, $Z$ and $B_\psi$, with $B_\theta$, $B_\alpha$, $\bar{\iota}$, the axis and $B$ as inputs. In this procedure, we observed that $Y$ and $B_\psi$ are determined up to a free $\chi$-independent function. Furthermore, at every other order, there is an ordinary differential equation on $Y_{n1}$ to solve for. We focus on this equation now.

\section{Generalized $\sigma$-equation and magnetic shear}

We shall refer to the recurrent ordinary differential equation for $Y_{n1}^C$ by the name of \textit{generalized $\sigma$-equation}. We do so because, to leading order, $C_\perp^1$ yields the well-known Riccati $\sigma$-equation.\cite{garrenboozer1991b,landreman2018a} The standard $\sigma$-equation (see Eq.~(26) in [\onlinecite{rodriguez2020i}]) reads,
\begin{equation}
    \frac{\mathrm{d}\sigma}{\mathrm{d}\phi}=-\Bar{\iota}_0\left[1+\frac{1}{4B_0}\left(\frac{\eta}{\kappa}\right)^4+\sigma^2\right]+\frac{B_{\alpha0}}{2}(2\tau+B_{\theta20})\left(\frac{\eta}{\kappa}\right)^2, \label{eqn:RiccatiSigma}
\end{equation}
where $\eta=-B_{11}^C/2B_0$ and $Y_{11}^C=\sigma Y_{11}^S$. The existence of solutions of this non-linear ODE has been thoroughly analysed in [\onlinecite{landreman2019}]. Importantly, the authors proved in Appendix A of their work that there exists a unique periodic solution pair $\{\sigma,\bar{\iota}_0\}$ for given $\{\eta,\sigma(0),B_{\theta20},\kappa,\tau\}$. The important lesson learned is that the equation describes not only $\sigma$ (and thus provide $Y_{11}^C$ as a solution) but also constraints the rotational transform on axis $\iota_0$. The value of $\bar{\iota}_0$ is strongly constrained to guarantee that the solutions to the equation are periodic in $\phi$. \par
With this leading order in mind, we explore the higher-order form of the $\sigma$-equation. Let us start by obtaining a closed form for the differential equation. Looking at the $m=0$ component of Eq~(\ref{eq:alterC2}), the equation for $B_\theta$, and focusing on terms including $Y_n$, we write
\begin{multline}
    B_{\theta n+1,0}B_{\alpha0}B_0=\left[\tau l'Y_n\partial_\chi X_1+\partial_\chi Y_n\left(\partial_\phi Y_1-X_1\tau l'+\right.\right.\\
    \left.\left.+\bar{\iota}_0\partial_\chi Y_1\right)+\partial_\chi Y_1\left(\partial_\phi Y_n+\bar{\iota}_0\partial_\chi Y_n\right)\right]_{m=0}+\Xi_n,
\end{multline}
where $\Xi_n$ denotes terms in the equation that do not potentially include $Y_n$ nor $Z_{n+1}$, and $l'$ is shorthand for $\mathrm{d}l/\mathrm{d}\phi$. The subscript $m=0$ indicates that only the $m=0$ harmonic of the square bracket should be picked. Because the differential equation on $Y_{n1}^C$ only occurs at odd $n$, it follows that
\begin{gather*}
    \left(Y_n X_1\right)_{m=0}=\frac{1}{2}X_{11}^CY_{n1}^C, \\
    \left(Y_nY_1\right)_{m=0}=\frac{1}{2}Y_{11}^S\left(Y_{n1}^C\sigma+Y_{n1}^S\right).
\end{gather*}
With this, and conveniently defining $Y_{n1}^C=Y_{11}^S\tilde{\sigma}_n$, we may collect terms to write the following succinct linear ODE,
\begin{equation}
    \frac{\mathrm{d}\tilde{\sigma}_n}{\mathrm{d}\phi}+2\bar{\iota}_0\sigma\tilde{\sigma}_n=\Lambda_n+\frac{2B_{\alpha0}B_0}{\left(Y_{11}^S\right)^2}B_{\theta n+1,0}, \label{eqn:genSigEq}
\end{equation}
where,
\begin{multline*}
    \Lambda_n=\frac{(Y_{n1}^S)'}{Y_{11}^S}\sigma-2\bar{\iota}_0\frac{Y_{n1}^S}{Y_{11}^S}-\frac{Y_{n1}^S(Y_{11}^C)'}{(Y_{11}^S)^2}+\\
    +2\tau l'\frac{Y_{n1}^SX_{11}^C}{(Y_{11}^S)^2}-2\frac{\Xi_n}{(Y_{11}^S)^2}.
\end{multline*}
We call this differential equation (\ref{eqn:genSigEq}) the \textit{generalized $\sigma$ equation}. It is the generalization of the Riccati equation for $\sigma$, Eq.~(\ref{eqn:RiccatiSigma}), and has the same structure at all odd orders. In the form it is written, the equation is a linear ODE for $\tilde{\sigma}_n$. We repeat for emphasis that this equation is \textit{linear}, unlike the Riccati equation. \par
The linearity of the ODE facilitates considerations of the existence of its solution. We can find a closed form for the solution of Eq.~(\ref{eqn:genSigEq}) using an integration factor,
\begin{multline}
    \tilde{\sigma}_n=e^{-2\bar{\iota}_0\int_0^\phi\sigma\mathrm{d}\phi'}\left\{\tilde{\sigma}_n(0)+\frac{}{}\right.\\
    \left.+\int_{0}^\phi\mathrm{d}\phi' e^{2\bar{\iota}_0\int_0^{\phi'}\sigma\mathrm{d}\phi''}\left[\Lambda_n+\frac{2B_{\alpha0}B_0}{(Y_{11}^S)^2}B_{\theta n+1,0}\right]\right\}, \label{eqn:perCondSigEq}
\end{multline}
where we have explicitly introduced the `initial' condition $\tilde{\sigma}_n(0)$. 
\par
Although we have successfully written a closed-form solution to the generalized $\sigma$ equation, this solution is not yet guaranteed to be physically realizable in a torus. The function $\tilde{\sigma}_n$, which describes some aspect of the shape of the flux surfaces, must be periodic in $\phi$. \par
To impose periodicity, we capitalize on the periodicity of the generalized $\sigma$-equation (\ref{eqn:genSigEq}). All coefficients in the equation, including the inhomogeneous term, are periodic. This guarantees that, as the ODE is first order, the solution is periodic iff $\tilde{\sigma}_n(0)=\tilde{\sigma}_n(2\pi)$. Using the closed form of the solution, the periodicity requirement may be then given as
\begin{multline}
    \int_{0}^{2\pi}\mathrm{d}\phi' e^{2\bar{\iota}_0\int_0^{\phi'}\sigma\mathrm{d}\phi''}\left[\Lambda_n+\frac{2B_{\alpha0}B_0}{(Y_{11}^S)^2}B_{\theta n+1,0}\right]=\\
    =\tilde{\sigma}_n(0)\left(e^{2\bar{\iota}_0\int_0^{2\pi}\sigma\mathrm{d}\phi'}-1\right). \label{eqn:perCondGenSig}
\end{multline}
If the input parameters are not chosen appropriately, this condition is generally not satisfied. We now explore the conditions imposed by (\ref{eqn:perCondGenSig}) and their physical interpretation.\footnote{It is customary to assume stellarator symmetry, in which case $\sigma$ is an odd function in $\phi$, and the periodicity condition in Eq.~(\ref{eqn:perCondGenSig}) becomes independent of $\tilde{\sigma}_n(0)$.} \par
\paragraph{Rotational transform.} We propose first to consider the constraint equation (\ref{eqn:perCondGenSig}) as an equation on the average toroidal current $\bar{B}_{\theta n0}=\int_0^{2\pi}B_{\theta n0}\mathrm{d}\phi$. Reorganizing terms and defining $B_{\theta n+1,0}=\bar{B}_{\theta n+1,0}+\tilde{B}_{\theta n+1,0}$, we obtain 
\begin{multline}
    \bar{B}_{\theta n+1,0}=-\frac{\int_{0}^{2\pi}\mathrm{d}\phi' e^{2\bar{\iota}_0\int_0^{\phi'}\sigma\mathrm{d}\phi''}\left[\Lambda_n+\frac{2B_{\alpha0}B_0}{(Y_{11}^S)^2}\tilde{B}_{\theta n+1,0}\right]}{\int_0^{2\pi}\frac{2B_{\alpha 0}B_0}{(Y_{11}^S)^2}e^{2\bar{\iota}_0\int_0^{\phi'}\sigma\mathrm{d}\phi''}}+\\
    +\tilde{\sigma}_n(0)\frac{e^{2\bar{\iota}_0\int_0^{2\pi}\sigma\mathrm{d}\phi'}-1}{\int_0^{2\pi}\frac{2B_{\alpha 0}B_0}{(Y_{11}^S)^2}e^{2\bar{\iota}_0\int_0^{\phi'}\sigma\mathrm{d}\phi''}}. \label{eqn:currSelfCons}
\end{multline}
Thus, by choosing the average toroidal current profile as in Eq.~(\ref{eqn:currSelfCons}), the problem can be tailored to support solutions to the generalized $\sigma$ equation to any arbitrarily high order. This formal construction is physically intuitive. The average toroidal current must be prescribed carefully to achieve a particular form of the rotational transform (considered an input in this context). This particular choice of current affects the average but leaves the angle-dependent part of $B_\theta$ free (the average current would in Boozer coordinates correspond to the flux function $I(\psi)$). The idea is, in essence, no different from the axisymmetric Grad-Shafranov problem, where prescribing the rotational transform leaves the toroidal current to be self-consistently determined. \par
Physically, one expects to find that an appropriate current profile can always be chosen to achieve the required rotational transform profile. This is the insight formally encoded in the construction of Eq.~(\ref{eqn:currSelfCons}). Of course, it is not enough to provide the construction explicitly as we have, but we may also show that the construction is well-behaved. The denominator on the right-hand side is non-zero to all orders, as $Y_{11}^S\neq 0$ and the exponential is always positive. Hence, our physical intuition is formally justified. \par
The choice of rotational transform as an input to the problem is not the standard approach taken for the leading Riccati equation. There it was the current density on-axis ($B_{\theta20}$), which was given as an input for the rotational transform to be evaluated. Of course, these are two sides of the same coin. This duality can be seen in practice when solving 3D magnetic MHD equations. It is possible to provide either the rotational transform or the toroidal current profile as inputs, as it occurs in practical, numerical solvers such as VMEC or SPEC. {  See Appendix E for further considerations on the well-posedness of the problem in this reverse form.} \par
{   Before moving on, we mention that it might be possible to avoid having to solve Eq.~(\ref{eqn:genSigEq}) altogether if one treats $\tilde{\sigma}_n$ rather than $B_{\theta n0}$ as an input. Although formally this is a well-posed problem, it is more physically convenient to deal with the problem in the form proposed. Having freedom in the choice of current allows for a better coupling to equilibria conditions. For instance, to require that the toroidal current is a flux function. } 

\paragraph{Toroidal current.} Let us now take the alternative point of view on the solvability condition (\ref{eqn:perCondGenSig}). Take the toroidal current profile to be prescribed as an input. From a practical perspective, this might be the most natural formulation, as one may exercise more control over the currents driven in a device. The rotational transform sets itself as a result.  \par
To obtain this relation explicitly, we need to collect terms in $\Lambda_n$ that include explicitly $\bar{\iota}_n$. This can be done by looking at the $B_\theta$ equation (\ref{eq:alterC2}). To leading order, the highest $\bar{\iota}$ terms may only come from those terms explicit in $\bar{\iota}$. For the $n$-th order, we pick $\bar{\iota}_{(n-1)/2}$ terms. All the other functions $X,~Y$ and $Z$, to their $n-1$ (or lower) orders do not include this high-order component of the rotational transform. Then we may formally write,
\begin{equation}
    \Lambda_n=\tilde{\Lambda}_n-\bar{\iota}_{(n-1)/2}\left[1+\sigma^2+\frac{1}{4B_0}\left(\frac{\eta}{\kappa}\right)^4\right].
\end{equation}
Importantly, the term in brackets is a positive non-zero function. Its form is reminiscent of the Riccati $\sigma$ equation (\ref{eqn:RiccatiSigma}). (In fact, it is the piece that is multiplied by $\bar{\iota}_0$ in (\ref{eqn:RiccatiSigma})). At every order, the non-zero value of this factor makes it possible to interpret the solvability condition as a constraint on the rotational transform profile. The solution $\tilde{\sigma}_n$ is periodic iff,
\begin{multline}
    \bar{\iota}_{(n-1)/2}=\frac{\int_0^{2\pi}\mathrm{d}\phi' e^{2\bar{\iota}_0\int_0^{\phi'}\sigma\mathrm{d}\phi''}\left[\tilde{\Lambda}_{n}+\frac{2B_{\alpha0}B_0}{\left(Y_{11}^S\right)^2}B_{\theta n+1,0}\right]}{\int_0^{2\pi}\mathrm{d}\phi'e^{2\bar{\iota}_0\int_0^{\phi'}\sigma\mathrm{d}\phi''}\left[1+\sigma^2+\frac{1}{4B_0}\left(\frac{\eta}{\kappa}\right)^4\right]}+\\
    +\tilde{\sigma}_{n}(0)\frac{1-e^{2\bar{\iota}_0\int_0^{2\pi}\sigma\mathrm{d}\phi'}}{\int_0^{2\pi}\mathrm{d}\phi'e^{2\bar{\iota}_0\int_0^{\phi'}\sigma\mathrm{d}\phi''}\left[1+\sigma^2+\frac{1}{4B_0}\left  (\frac{\eta}{\kappa}\right)^4\right]}. \label{eqn:condIotaGenSigma}
\end{multline}
Thus, a periodic solution to the generalized $\sigma$-equation exists provided that the rotational transform profile is consistently chosen. By requiring this for all odd $n$, the rotational transform profile is determined as a power series in $\epsilon$ order by order. \par
The first-order term in the rotational transform series is of special importance. This parameter, $\bar{\iota}_1$, corresponds to the magnetic shear. The procedure presented above offers a way to construct the magnetic shear in the context of near-axis expansions. Both the current (quite explicitly in (\ref{eqn:condIotaGenSigma})) and the geometry (through all the lower order $X$, $Y$ and $Z$ pieces) influence the shear. The convoluted nature of the expansion at third order (where the magnetic shear shows up) complicates a more detailed analysis of the magnetic properties that affect the shear. A direct construction approach following [\onlinecite{jorge2020}] would likely offer more illuminating expressions from the geometric perspective and complement the inverse approach. This is the case for the rotational transform on axis, which in the direct representation has a succinct form\cite{Helander2014,jorge2020}. \par
Although convoluted, we have presented a procedure by which one may obtain the magnetic shear explicitly. As indicated, the shear appears in third order in the expansion. However, note that its construction follows from the solvability condition of the generalized $\sigma$-equation. Thus, it includes information concerning lower orders. This observation means we do not need all the information available in the third order to compute the shear. A thoughtful approach can therefore spare unnecessary labor. Considering the carefully described dependencies in [\onlinecite{rodriguez2020i}] and the constructions discussed in the appendices, one can show that knowledge of the following third-order functions is all that is required: $Z_{31}^C$, $Z_{31}^S$, $X_{31}^C$, $X_{31}^S$ and $Y_{31}^S$, for which closed forms are given in Appendix F. This also requires specifying $B_{31}^C$ and $B_{31}^S$, as well as using the procedure in Appendix D to find $B_{\psi 11}$. Together with the lower-order solutions, this is all that is needed to compute the magnetic shear. \par
Although this procedure is presented for a weakly quasisymmetric field, the appearance of shear as a solvability condition in the near axis construction prevails even if QS conditions are relaxed. An analogous generalized $\sigma$ equation and solvability condition are encountered in a general 3D field scenario in an ideal (non-QS) MHD balance. In that case, the difference is on the $\phi$ dependence of the $B$ coefficients. Nevertheless, the formal origin of the shear would be analogous. {  In this context, the choice of $\tilde{\sigma}_n$ as the unknown of Eq.~(\ref{eqn:genSigEq}) is important.} Another potentially interesting scenario would be that of weak QS with the addition of some form of force balance ( [\onlinecite{rodriguez2020i}]). The requirements on the generalized $\sigma$ equation can be used to compute shear consistently in those scenarios as well. \par

\section {General structure of the construction of solutions}
We have shown constructively that weakly quasisymmetric solutions may be constructed by near-axis expansion to arbitrarily high order with the given prescription. The construction of solutions only requires inverting non-singular algebraic operators (explicitly solved in the Appendices) and the solution to a recurrent ODE, the generalized $\sigma$-equation. The latter can always be found provided the rotational transform profile (or toroidal current) is consistently chosen. \par
This systematic solution of the magnetic equations is essential at two levels. For one, it shows that the construction of weakly quasisymmetric fields is a priori not limited by the Garren-Boozer overdetermination conundrum\cite{garrenboozer1991b}. This appears to overturn the conventional belief on the limitation of global QS solutions from a fluid perspective. Of course, showing that every order in the expansion can be solved is not enough to claim that there is a global solution. The character of the expansion has not yet been explored. In the generic case, the near-axis expansion could well correspond to a convergent (unlikely) or a {  divergent} asymptotic expansion (more likely). {  Asymptotic in this context means that the truncated series at $N$ leaves a remainder of $O(\epsilon^{N+1})$. By the finite nature of all expansion coefficients shown in this paper, the expansion is so.} While a rigorous consideration of remainders requires further work, the systematic order-by-order construction presented in this paper provides a foundation (at least numerically) to explore the question. Since the expansion could be systematically carried out to arbitrarily high order, one could provide insight into how the magnitudes of the terms at different orders compare with each other. \par
\begin{figure*}
    \centering
    \includegraphics[width=0.9\textwidth]{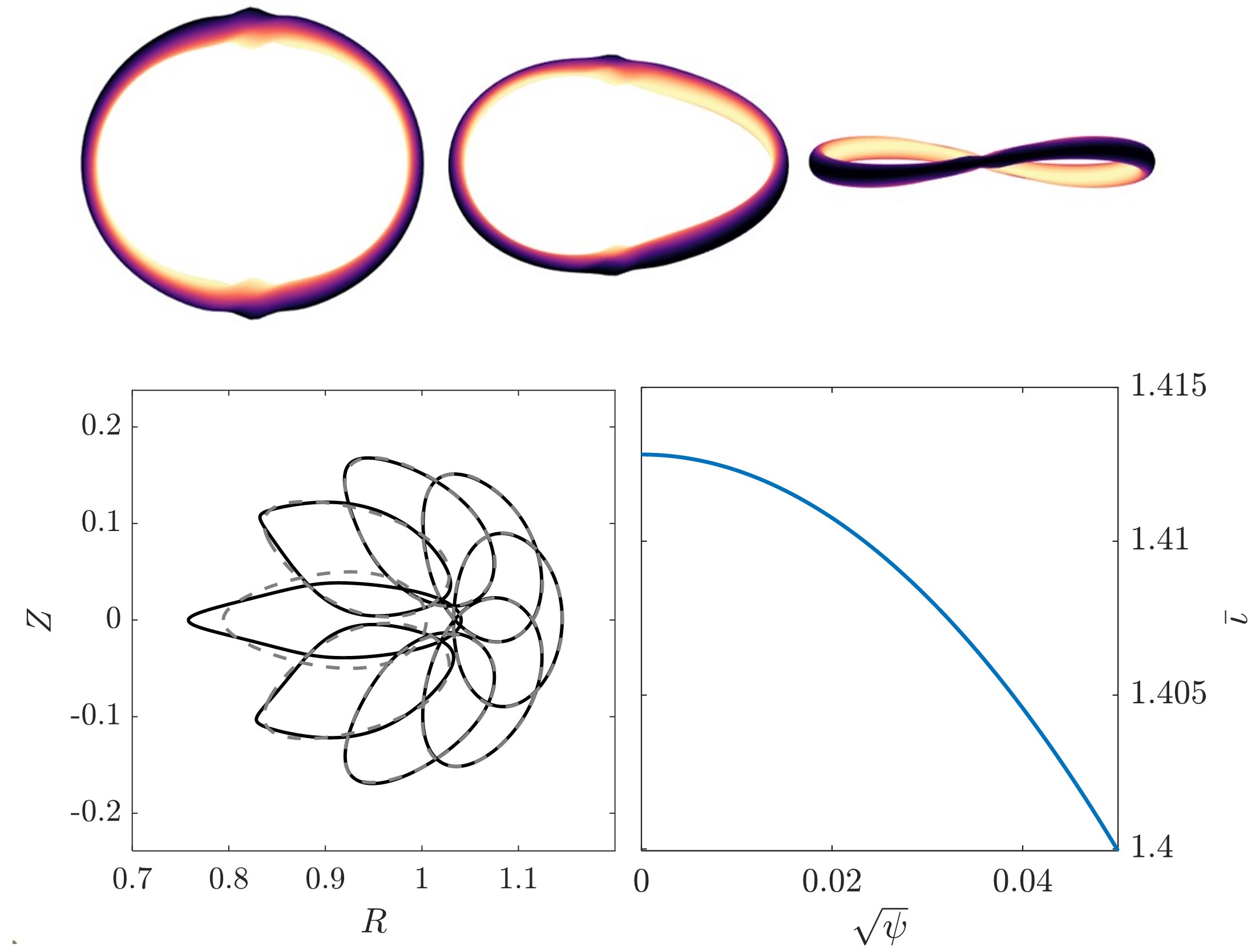}
    \caption{\textbf{Weakly quasisymmetric configuration to third order.} The figure shows aspects of a quasisymmetric configuration in which the first three terms in the expansion are included. The upper figures are different projections of a 3D rendering of the flux surface at $\epsilon=0.05$ where the color corresponds to the magnetic field magnitude. The lower left plot shows the cross-sections of the configuration at different values of the cylindrical angle (broken lines showing the contours at second-order). The lower right shows the predicted leading rotational transform profile. The configuration is an extension on the example presented in [\onlinecite{rodriguez2021pre}]:  $\sigma(0)=1.01\times10^{-4}$, $\tilde{\sigma}_3(0)=0$, $\bar{B}_{\theta 20}=2.8546$, $B_{\theta 31}=0$, $B_{\theta 40}=0$, $\eta/\sqrt{2}=0.95$, $p_0=0.08$, $\Delta_0=0$, $B_{22}^C=5.51$, $B_{22}^S=0$, $B_{20}=-3.69$, $B_{31}^C=0.01$, $B_{31}^S=0.01$, $R_\mathrm{ax}=1+0.09\cos2\phi$, $Z_\mathrm{ax}=-0.09\sin2\phi$, $B_{\alpha0}=1.02$, $B_{\alpha1}=2.04$, $\epsilon=0.05$ (aspect ratio of $20$).}
    \label{fig:IOnaeMagEq}
\end{figure*}
{  As a way of example, we present a numerical solution that is exactly quasisymmetric up to and including third order. We do so by extending the solution in [\onlinecite{rodriguez2021pre}], which to second-order corresponds to a QS field in equilibrium with anisotropic pressure. We extend it by relaxing force balance at third order. The third-order shaping forces the configuration to a larger aspect ratio to avoid unphysical self-intersection of flux surfaces. This could be taken as indicative of a potentially diverging series. It is also remarkable that most of the shaping contributions pile up in the $\phi=\pi$ Section of the configuration. This paper opens the door to investigating these matters and their potential implications.}
\par
The second important aspect consolidated in this paper is the input/output structure of the problem. We have provided a complete and \textit{consistent} constructive approach to the solution of weak QS. In particular, we have shown the most convenient form for formulating the relevant equations, explicitly identifying those functions that need to be provided as an input and those that come out of the constraint equations. A schematic of this is given in Fig.~\ref{fig:IOnaeMagEq}. \par
\begin{figure*}
    \centering
    \includegraphics[width=0.5\textwidth]{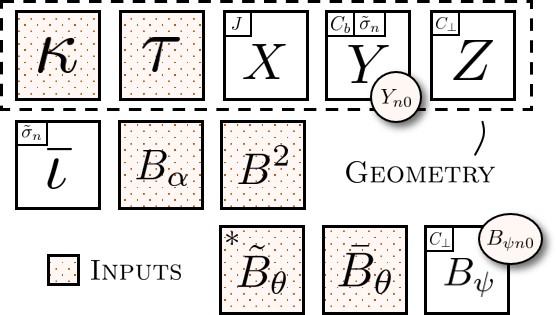}
    \caption{\textbf{Schematic description of the input/output functions for the magnetic equations.} The diagram shows the functions involved in describing the magnetic field configuration. Shaded in red are the functions that are given as inputs to the construction. The asterisk in $\tilde{B}_\theta$ indicates that this piece is not completely free, as it must satisfy $B_{\theta n n}=0$ from $C_\perp$. The upper left corners in the output functions indicate the equation from which these functions are computed. Finally, the round details indicate freedom in choosing a piece of $Y$ and $B_\psi$. The top row constitutes all functions describing the geometry of the magnetic flux surfaces.}
    \label{fig:IOnaeMagEq}
\end{figure*}
Following the prescription detailed in the diagram of Fig.~\ref{fig:IOnaeMagEq} the weak QS problem is well-posed at the level of perturbation theory. The field will eventually have to be considered in some form of force balance for the practical implementation of a quasisymmetric configuration. The form of equilibrium will impose additional constraints on the form of the magnetic configuration. As a result, some of those functions that appeared free in the context of the magnetic equations will now be subject to additional constraints. Depending on the form of the forcing, these constraints may or may not be reconcilable with the requirements from the magnetic equations. Examples of expansions of weak QS in equilibria are isotropic MHD equilibria\cite{garrenboozer1991b,landreman2018a} and anisotropic pressure equilibria\cite{rodriguez2020i,rodriguez2020ii}. The completeness in the magnetic equation description presented has value, as it constitutes a sub-problem of the larger problem.
%%%%%%%%%%%%%%%%%%%%%%%%%%%%%%%%%%%%%%%%%%%%%%%%%%%%

\section{Conclusion}
In this paper, we prove that near-axis expansions of weakly quasisymmetric fields can be performed to arbitrarily high order when disregarding the underlying form of equilibria. This suggests a possible pathway for the construction of global quasisymmetric solutions (at least numerically). \par
By virtue of the expansion, most of the governing constraint equations are algebraic equations that can be solved explicitly. We present, drawing directly and extending on the work in [\onlinecite{rodriguez2020i}], closed-form expressions for the unknowns. At every other order, an ordinary differential equation needs to be solved.
We call these linear ODEs generalized $\sigma$-equations and present closed-form solutions. For periodic solutions to exist at all orders, expansion parameters need to be chosen appropriately. In particular, either the average toroidal current or rotational transform profiles must be appropriately tailored, taking either of them as outputs of the problem to be determined self-consistently. This gives a straightforward (though algebraically demanding) way of computing magnetic shear in the near-axis framework to guarantee the periodicity of solutions. The construction of magnetic shear in the near axis expansion can be straightforwardly extended to more general (non-QS) fields.  \par
The constructive proof presented here can be applied to systematically solve the near-axis expansion to arbitrarily high order with finite expansion coefficients. An implementation to study issues such as convergence or asymptotic properties { (including truncation)} is left for future work. {  We present a numerical example to third order.} However, the constructive structure of the solution presented provides insight into the appropriate way to formulate the quasi-symmetric equilibrium problem and the appropriate choice of free functions.

\appendix
\section{Constructing $X_n$}
The function $X_n$ can be obtained in closed form from the Jacobian equation. To that purpose, we write the explicit form of equation $J$ (Eq.~(12) in [\onlinecite{rodriguez2020i}]),
\begin{align}
    \frac{B_\alpha^2}{B^2}=\left(\Bar{\iota}\partial_\chi X+\partial_\phi X+\tau Y\frac{\mathrm{d}l}{\mathrm{d}\phi}+Z\kappa\frac{\mathrm{d}l}{\mathrm{d}\phi}\right)^2+\nonumber\\
    +\left(\Bar{\iota}\partial_\chi Y+\partial_\phi Y-X\tau\frac{\mathrm{d}l}{\mathrm{d}\phi}\right)^2+\nonumber\\
    +\left(\Bar{\iota}\partial_\chi Z+\partial_\phi Z-X\kappa\frac{\mathrm{d}l}{\mathrm{d}\phi}+\frac{\mathrm{d}l}{\mathrm{d}\phi}\right)^2. \label{eq:Jgen}
\end{align}
To order $\epsilon^n$, the only possible $X_n$ term comes from the last squared bracket and, in particular, the cross-product between the last two terms in the bracket. Thus, to order $\epsilon^n$ we obtain,
\begin{equation}
    -2X_n\kappa\left(\frac{\mathrm{d}l}{\mathrm{d}\phi}\right)^2=J^n,
\end{equation}
where $J^n$ denotes all remaining terms from the equation to order $\epsilon^n$. This gives a systematic, well-behaved form of constructing $X_n$ to arbitrarily high order. This is so because for a quasisymmetric system, $\mathrm{d}l/\mathrm{d}\phi\neq 0$ is a non-zero constant, and $\kappa$ must vanish nowhere { from $n=1$}. Doing so would lead to the leading-order elliptical cross-sections of flux surfaces near the axis to be infinitely elongated (ribbon-like) along the normal direction to the axis. 

\section{Algebraic construction $Y_n$}
The construction of $Y_n$ follows from $C_b$, the tangential projection of the co(ntra)variant equation. The equation at order $n$ has $n+1$ harmonic components, and thus $n+1$ independent constraint equations. These should be used to solve for $Y_{n+1}$ much in the fashion that the lower orders were shown in previous studies\cite{rodriguez2020i,rodriguez2020ii}. To show this, we shall consider $C_b^n$ and focus on the $Y_{n+1}$ coefficient.\par
First, we explicitly write $C_b$, (Eq.~(17) in [\onlinecite{rodriguez2020i}])
\begin{widetext}
\begin{align}
    &-(B_\alpha-\bar{\iota}B_\theta)\left(\partial_\chi X\partial_\psi Y-\partial_\psi X\partial_\chi Y\right)-B_\psi\left[\partial_\chi Y\left(\partial_\phi X+\tau Y\frac{\mathrm{d}l}{\mathrm{d}\phi}+Z\kappa\frac{\mathrm{d}l}{\mathrm{d}\phi}\right)-\partial_\chi X\left(\partial_\phi Y-X\tau\frac{\mathrm{d}l}{\mathrm{d}\phi}\right)\right]+\nonumber\\
    &+B_\theta\left[\partial_\psi Y\left(\partial_\phi X+\tau Y\frac{\mathrm{d}l}{\mathrm{d}\phi}+Z\kappa\frac{\mathrm{d}l}{\mathrm{d}\phi}\right)-\partial_\psi X\left(\partial_\phi Y-X\tau\frac{\mathrm{d}l}{\mathrm{d}\phi}\right)\right]=\left(\partial_\phi Z-X\kappa\frac{\mathrm{d}l}{\mathrm{d}\phi}+\frac{\mathrm{d}l}{\mathrm{d}\phi}\right)+\Bar{\iota}\partial_\chi Z, \label{eq:Cb}
\end{align}
\end{widetext}
we may write the relevant $O(\epsilon^n)$, $Y_{n+1}$ dependent pieces as, (see Eq.~(D6) in [\onlinecite{rodriguez2020i}])
\begin{equation}
    -\frac{B_{\alpha0}}{2}\left[\Bar{X}_1(n+1)Y_{n+1}-mX_1\Bar{Y}_{n+1,m}\right]=C_b^n, \label{eqn:solveYalg}
\end{equation}
where $C_b^n$ represent the remaining terms in the equation that do not depend on $Y_{n+1}$ explicitly, and the overbar represents derivative with respect to $\chi$. From the combination of terms above it might appear that there are in fact $n+2$ constraint equations associated to the $\chi$ harmonics of the equation (as the $Y_{n+1}$ piece presented seems to have $n+2$ components). However, due to the commutator-like structure of the term, the $n+2$ harmonic vanishes trivially. (This vanishing constraints $B_{\theta nn}=0$, as is well known and had been previously shown\cite{rodriguez2020i}.) \par
To show that there always exist non-singular solutions to $Y_{n+1}$, we present a systematic construction of $Y$ now. Recalling from the lowest orders (Eq.~(14)-(15) in [\onlinecite{rodriguez2020i}]) that $X_1=(\eta/\kappa)\cos\chi$ and capitalising on trigonometric relations of the form $\sin mx\sin x=1/2[\cos(m-1)x-\cos(m+1)x]$, we may explicitly construct $Y_n$, as follows:
\begin{itemize}
    \item for $n=2k,~k\in\mathbb{Z}$,
    \begin{subequations}
        \begin{align}
            Y_{n+1,1}^S=&\frac{4\kappa}{B_{\alpha0}\eta (n+2)}C_b^n|_{m=0},\\
            Y_{n+1,m+2}^S=&\frac{1}{n+m+3}\left[(n+1-m)Y_{n+1,m}^S+\frac{4\kappa}{B_{\alpha0}\eta}C_b^n|_{m+1}^C\right], \\
            Y_{n+1,m+2}^C=&\frac{1}{n+m+3}\left[(n+1-m)Y_{n+1,m}^C-\frac{4\kappa}{B_{\alpha0}\eta}C_b^n|_{m+1}^S\right].
        \end{align}
    \end{subequations}
    The latter two hold for $1\leq m\leq n-1$, and $m$ is odd, for which there is no singularity. The nomenclature $C_b^n|_m^{C/S}$ refers to the $m$-th cosine or sine harmonic of the equation excluding the $Y_{n+1}$ terms.
    \item for $n=2k-1,~k\in\mathbb{Z}$,
    \begin{subequations}
        \begin{align}
            Y_{n+1,2}^S=&\frac{4\kappa}{B_{\alpha0}\eta}\frac{1}{n+3}C_b^n|_{m=1}^C, \\
            Y_{n+1,2}^C=&\frac{1}{n+3}\left[2Y_{n+1,0}^C-\frac{4\kappa}{B_{\alpha0}\eta}C_b^n|_{m=1}^S\right], \\
            Y_{n+1,m+2}^C=&\frac{1}{n+m+3}\left[(n+1-m)Y_{n+1,m}^C-\frac{4\kappa}{\eta B_{\alpha0}}C_b^n|_{m+1}^S\right], \\
            Y_{n+1,m+2}^S=&\frac{1}{n+m+3}\left[\frac{4\kappa}{B_{\alpha0}\eta}C_b^n|_{m+1}^C-(m-n-1)Y_{n+1,m}^S\right].
        \end{align}
    \end{subequations}
    The latter two equations hold for $2\leq m\leq n-1$, for which there is no singularity. Recall, as always, that any function $f_{nm}=0$ if $m>n$, in addition to the parity being shared by $n$ and $m$.
\end{itemize}
\par
Thus, $n+1$ components of $Y_{n+1}$ can be uniquely found via algebraic considerations. It is also clear from the construction that $Y_{n0}$ and $Y_{n1}^C$ are not given by this construction. What happens to these remaining unconstrained functions? At every other order one of the $C_\perp$ constraints should be employed to constrain $Y$ in the form of a differential equation on $Y_{n 1}^C$ (see Section III for the details). The other one, however, remains an unconstrained function and will play a role when force balance is considered. This free function corresponds to the $Y_{n0}$ at every even order, representing the free addition of a $\chi$-independent function to the function $Y$.

\section{Constructing $Z$} \label{sec:app:constrZ}
In this Appendix we discuss the general procedure to obtain the form of $Z_n$ to arbitrary high order in the near-axis expansion. The general structure was presented in Appendix D of [\onlinecite{rodriguez2020i}], so here we reproduce the procedure sketched there, with an emphasis on the good behaviour of the algebraic inversion of the equation. \par
To obtain an explicit close form for $Z_{n+1}$ consider the order $\epsilon^n$ $C_\kappa$ and $C_\tau$ equations. For completeness, the $\hat{\kappa}_0$ and $\hat{\tau}_0$ components of the co(ntra)variant equation are, (Eqs.~(18)-(19) in [\onlinecite{rodriguez2020i}])
\begin{widetext}
\begin{multline}
    -(B_\alpha-\bar{\iota}B_\theta)\left(\partial_\chi Y\partial_\psi Z-\partial_\psi Y\partial_\chi Z\right)-B_\psi\left[\partial_\chi Z\left(\partial_\phi Y-X\tau\frac{\mathrm{d}l}{\mathrm{d}\phi}\right)-\partial_\chi Y\left(\partial_\phi Z-X\kappa\frac{\mathrm{d}l}{\mathrm{d}\phi}+\frac{\mathrm{d}l}{\mathrm{d}\phi}\right)\right]+\\
    +B_\theta\left[\partial_\psi Z\left(\partial_\phi Y-X\tau\frac{\mathrm{d}l}{\mathrm{d}\phi}\right)-\partial_\psi Y\left(\partial_\phi Z-X\kappa\frac{\mathrm{d}l}{\mathrm{d}\phi}+\frac{\mathrm{d}l}{\mathrm{d}\phi}\right)\right]=\left(\partial_\phi X+\tau Y\frac{\mathrm{d}l}{\mathrm{d}\phi}+Z\kappa\frac{\mathrm{d}l}{\mathrm{d}\phi}\right)+\Bar{\iota}\partial_\chi X, \label{eq:Ck}
\end{multline}
\begin{multline}
    -(B_\alpha-\bar{\iota}B_\theta)\left(\partial_\chi Z\partial_\psi X-\partial_\psi Z\partial_\chi X\right)-B_\psi\left[\partial_\chi X\left(\partial_\phi Z-X\kappa\frac{\mathrm{d}l}{\mathrm{d}\phi}+\frac{\mathrm{d}l}{\mathrm{d}\phi}\right)-\partial_\chi Z\left(\partial_\phi X+\tau Y\frac{\mathrm{d}l}{\mathrm{d}\phi}+Z\kappa\frac{\mathrm{d}l}{\mathrm{d}\phi}\right)\right]+\\
    +B_\theta\left[\partial_\psi X\left(\partial_\phi Z-X\kappa\frac{\mathrm{d}l}{\mathrm{d}\phi}+\frac{\mathrm{d}l}{\mathrm{d}\phi}\right)-\partial_\psi Z\left(\partial_\phi X+\tau Y\frac{\mathrm{d}l}{\mathrm{d}\phi}+Z\kappa\frac{\mathrm{d}l}{\mathrm{d}\phi}\right)\right]=\left(\partial_\phi Y-X\tau\frac{\mathrm{d}l}{\mathrm{d}\phi}\right)+\Bar{\iota}\partial_\chi Y. \label{eq:Ct}
\end{multline}
\end{widetext}
The leading $Z$ functions are actually one order higher than $\epsilon^n$, as they appear together with flux derivatives. Highlighting the relevant terms in $Z_{n+1}$, we obtain
\begin{gather*}
    -B_{\alpha0}\left[\Bar{Y}_1\frac{n+1}{2}Z_{n+1,m}-m\Bar{Z}_{n+1,m}\frac{1}{2}Y_1\right]=C_\kappa^n \\
    -B_{\alpha0}\left[-\Bar{X}_1\frac{n+1}{2}Z_{n+1,m}+m\Bar{Z}_{n+1,m}\frac{1}{2}X_1\right]=C_\tau^n,
\end{gather*} 
where the equation labels denote the remaining terms in the equations not including the highlighted $Z_{n+1}$ pieces. Capitalising on the leading order $C_b^0$ relation between $X_1$ and $Y_1$ (see Eq.~(23) in [\onlinecite{rodriguez2020i}]) that reads $B_{\alpha0}(X_1\partial_\chi Y_1-Y_1\partial_\chi X_1)=2\mathrm{d}l/\mathrm{d}\phi$, to obtain a closed form for $Z_{n+1}$ combine the equations above in the form $X_1 C_\kappa^n+Y_1 C_\tau^n$. Then,
\begin{equation}
    B_{\alpha0}\sqrt{B_0}(n+1)Z_{n+1,m}=X_1 C_\kappa^n+Y_1 C_\tau^n. \label{eqn:Zalg}
\end{equation}
The harmonics of $Z_{n+1}$ can then be read off this expression. Most importantly, the factors in front of $Z_{n+1}$, which need to be inverted to obtain a closed form for $Z$ are non-vanishing, implying that the algebraic equation will always yield well-behaved solutions for $Z$. \par

\section{Constructing $B_\psi$}
To obtain a closed form for $B_\psi$ let us consider an alternative form to $C_\perp$, by considering the projection along $\partial_\psi\mathbf{x}$ and $\partial_\chi \mathbf{x}$ of the co(ntra)variant equation. These forms are convenient in that they express the covariant magnetic field functions explicitly, (Eqs.~(D2)-(D3) in [\onlinecite{rodriguez2020i}])
\begin{widetext}
\begin{align}
    B_\psi J=&\partial_\psi X\left(\partial_\phi X+\tau Y\frac{\mathrm{d}l}{\mathrm{d}\phi}+Z\kappa\frac{\mathrm{d}l}{\mathrm{d}\phi}+\Bar{\iota}\partial_\chi X\right)+\partial_\psi Y\left(\partial_\phi Y-X\tau\frac{\mathrm{d}l}{\mathrm{d}\phi}+\Bar{\iota}\partial_\chi Y\right)+\nonumber\\
    &+\partial_\psi Z\left(\partial_\phi Z-X\kappa\frac{\mathrm{d}l}{\mathrm{d}\phi}+\frac{\mathrm{d}l}{\mathrm{d}\phi}+\Bar{\iota}\partial_\chi Z\right) \label{eq:alterC1}
\end{align}
\begin{align}
    B_\theta J=&\partial_\chi X\left(\partial_\phi X+\tau Y\frac{\mathrm{d}l}{\mathrm{d}\phi}+Z\kappa\frac{\mathrm{d}l}{\mathrm{d}\phi}+\Bar{\iota}\partial_\chi X\right)+\partial_\chi Y\left(\partial_\phi Y-X\tau\frac{\mathrm{d}l}{\mathrm{d}\phi}+\Bar{\iota}\partial_\chi Y\right)+\nonumber\\
   & +\partial_\chi Z\left(\partial_\phi Z-X\kappa\frac{\mathrm{d}l}{\mathrm{d}\phi}+\frac{\mathrm{d}l}{\mathrm{d}\phi}+\Bar{\iota}\partial_\chi Z\right). \label{eq:alterC2}
\end{align}
\end{widetext}
We have already seen (see Appendix \ref{sec:app:constrZ}) that the $C_\perp$ components of the co(ntra)variant equation had been used partially in obtaining closed forms for $Z$. Now we would like to exploit the remaining $n$ degrees of freedom left in the equation to construct $B_\psi$. To do so, we must eliminate $Z_{n+1}$ from Eqs.~(\ref{eq:alterC1}) and (\ref{eq:alterC2}). Looking at the leading order forms of the equations (order $\epsilon^{n-1}$ for the $B_\psi$ and $\epsilon^{n+1}$ for the $B_\theta$), the relevant terms can be expressed in the form,
\begin{gather*}
    B_{\theta n+1,m}B_{\alpha0}B_0=m\frac{\mathrm{d}l}{\mathrm{d}\phi}\Bar{Z}_{n+1,m}+C_{B_\theta}^{n+1} \\
    B_{\psi n-1,m}B_{\alpha0}B_0=\frac{n+1}{2}\frac{\mathrm{d}l}{\mathrm{d}\phi}Z_{n+1,m}+C_{B_\psi}^{n-1},
\end{gather*}
where the shorthand $C_{B_\theta}^{n+1}$ and $C_{B_\psi}^{n-1}$ refer to all remaining terms to the appropriate order in equaations (\ref{eq:alterC2}) and (\ref{eq:alterC1}) respectively. Combining these to eliminate $Z_{n+1}$, we have
\begin{gather}
    B_{\psi n-1,m}^C=C_{B_\psi}^{n-1}|_m^C-\frac{n+1}{2m}\left(B_{\theta n+1,m}^S-\frac{C_{B_\theta}^{n+1}|_m^S}{B_{\alpha0}B_0}\right), \label{eq:CpBpSBtC}\\
    B_{\psi n-1,m}^S=C_{B_\psi}^{n-1}|_m^S+\frac{n+1}{2m}\left(B_{\theta n+1,m}^C-\frac{C_{B_\theta}^{n+1}|_m^C}{B_{\alpha0}B_0}\right), \label{eq:CpBpCBtS}
\end{gather}
with $m=n-1,n-3\dots\in\mathbb{N}$. These constitute $n$ expressions for $B_{\psi, n-1}$ in terms of known functions (including the input-taken $B_\theta$). \par
The algebraic, systematic construction fails when $m=0$. This can be traced back to the original equations. When $m=0$, the $C_{B_\theta}$ equation corresponds precisely to the differential equation for $Y_{n 1}^C$ that we highlighted in the $Y$ construction considerations. This is indeed the generalized $\sigma$-equation dealt with in more detail in Section III. The $C_{B_\psi}$ equation for $m=0$ is nothing but an equation for $Z_{n+1,m}$ which we had already dealt with. This consistently leaves $B_{\psi n 0}$ undetermined. This explicit construction shows that the systematic construction of $B_\psi$ is consistent and constraints it up to a free $\chi$ independent function. 

\section{Solutions to the reverse problem $\sigma$ Riccati equation}
{  We consider in this Appendix the well-posedness of the problem of finding periodic solutions to the non-linear $\sigma$ Riccati equation (\ref{eqn:RiccatiSigma}) when the rotational transform is given as an input. A proof of existence and uniqueness of solutions exists for the regular formulation explored in [\onlinecite{landreman2019}]. It can be shown that if we specify $\sigma(0)$ and the current (as well as the axis shape and the field magnitude), there exists a unique pair $\{\sigma(\phi),\Bar{\iota}_0\}$ that solves the Riccati equation and is both bounded and periodic. \par
Such a proof does not exist for the reverse problem in which it is the on-axis rotational transform that is specified and the required amount of current sought as part of the solution. Physically, one expects to be able to regulate the current smoothly, affecting the rotational transform accordingly. We now show this to be the case, setting ourselves to do so using the existence and uniqueness of solutions in the regular form of the problem. \par
Let us start by abstractly writing the Riccati equation as,
\begin{equation}
    \frac{\mathrm{d}\sigma}{\mathrm{d}\phi}+\Bar{\iota}(P+\sigma^2)-(\Bar{Q}+\tilde{Q})=0, \label{eqn:riccatiSigmaQ}
\end{equation}
where we shall think of $\Bar{Q}$ as a constant offset that is directly related to the average toroidal current as $\Bar{Q}=\int\left(B_{\alpha0}(2\tau+B_{\theta20})\eta^2/\kappa^2\right)\mathrm{d}\phi/4\pi$ and $P=1+(\eta/\kappa)^4/4B_0$. The first step is to understand how solutions change when one makes changes $\delta\Bar{Q}$ in Eq.~(\ref{eqn:riccatiSigmaQ}). \par
Consider then two solution pairs $\{\sigma_1,\Bar{Q}_1\}$ and $\{\sigma_0,\Bar{Q}_0\}$, with $\bar{\iota}$, $\sigma(0)$ and all other elements being shared. Taking then the difference between the two Riccati equations, we obtain,
\begin{gather*}
    \frac{\mathrm{d}}{\mathrm{d}\phi}\left(\sigma_1-\sigma_0\right)+\Bar{\iota}(\sigma_1^2-\sigma_0^2)-(\Bar{Q}_1-\Bar{Q}_0)=0, \\
    \sigma_1-\sigma_0=(\Bar{Q}_1-\Bar{Q}_0)\underbrace{\left[e^{-\Bar{\iota}\int(\sigma_0+\sigma_1)\mathrm{d}\phi}\int_0^\phi e^{\Bar{\iota}\int(\sigma_0+\sigma_1)\mathrm{d}\phi'}\mathrm{d}\phi\right]}_{F(\phi)>0}.
\end{gather*}
Thus $\sigma$ is a \textit{monotonically increasing} function of $\Bar{Q}$. In particular, defining the periodicity measure\cite{landreman2019} $\Delta=\sigma(2\pi)-\sigma(0)$, $\Delta(\Bar{Q}+\delta\Bar{Q})-\Delta(\Bar{Q})=\delta \Bar{Q}F(2\pi)$. \par
We will now use the existence and uniqueness of solutions for the standard formulation of the Riccati $\sigma$ equation to argue uniqueness and existence of solutions in the rotational-transform mode of the equation. Start with a periodic solution $\Delta(\Bar{Q},\Bar{\iota})=0$ for some given values of rotational transform and $\Bar{Q}$. For any starting $\Bar{Q}$ there exists such a solution. Let us now consider nearby solutions to the one obtained. To obtain such solutions, introduce small variations in $\Bar{Q}$ and $\Bar{\iota}$ and require
\begin{equation}
    \Delta(\Bar{Q}+\delta\Bar{Q},\Bar{\iota}+\delta\Bar{\iota})\approx\frac{\delta\Delta}{\delta\Bar{Q}}\delta\Bar{Q} + \frac{\delta\Delta}{\delta\Bar{\iota}}\delta\Bar{\iota}=F(2\pi)\delta\Bar{Q}-\mathcal{F}(2\pi)\delta\Bar{\iota}\stackrel{!}{=}0.
\end{equation}
Here $\mathcal{F}$ is a positive function, defined in [\onlinecite{landreman2019}], which measures the variation in $\Delta$ with $\bar{\iota}$. In the limit of an infinitesimal variation (and assuming smoothness),
\begin{equation}
    \mathcal{F}=e^{-2\Bar{\iota}\int_0^\phi\sigma\mathrm{d}\phi'}\int_0^\phi e^{2\Bar{\iota}\int_0^{\phi'}\sigma\mathrm{d}\phi''}(P+\sigma^2)\mathrm{d}\phi'.
\end{equation}
With the infinitesimal limit of $F$ given as,
\begin{equation}
    F=e^{-2\Bar{\iota}\int_0^\phi\sigma\mathrm{d}\phi'}\int_0^\phi e^{2\Bar{\iota}\int_0^{\phi'}\sigma\mathrm{d}\phi''}\mathrm{d}\phi',
\end{equation}
it follows that
\begin{equation}
    0<\frac{\delta\Bar{\iota}}{\delta\Bar{Q}}=\frac{F(2\pi)}{\mathcal{F}(2\pi)}\leq1. \label{eqn:variationRicc}
\end{equation}
The bounds follow from $F,\mathcal{F}>0$ and the form of the functions $F$ and $\mathcal{F}$ themselves (in particular $P+\sigma^2\geq1$). Note that these functions can be understood as a result of linearising the equations around a known solution. Thus, we are assuming that some smoothness exists. \par
Eq.~(\ref{eqn:variationRicc}) gives the slope of a line of values $\{\bar{Q},\bar{\iota}\}$ for which the $\sigma$ solutions are periodic. In order for the reverse formulation to be well-posed (lead to uniqueness and existence), we require the curve to span the whole $\bar{\iota}$ range and be single-valued. \par
The variation $\delta\bar{\iota}/\delta\bar{Q}$ being monotonic is a good start for uniqueness. However, this condition is not sufficient for existence, as the solution might asymptote to a constant $\Bar{\iota}$ value for large $|\Bar{Q}|$. We shall proceed by reductio ad absurdum to show that this is not a possibility. We start with a known solution. From the bounded nature of $\sigma$ we may find the maximum of $\sigma$ at $\mathrm{d}\sigma/\mathrm{d}\phi=0$,
\begin{equation}
    P+\sigma^2\leq \frac{Q_\mathrm{max}}{\Bar{\iota}}.
\end{equation}
Assume that an asymptote in $\bar{\iota}$ exists, and look at the limit of large $\bar{Q}$. We may then take $Q_\mathrm{max}\propto\Bar{Q}$ and the lower bound of the gradient,
\begin{equation}
    \frac{\delta\Bar{\iota}}{\delta\Bar{Q}}\propto \frac{1}{\Bar{Q}}.
\end{equation}
This behavior is inconsistent with the existence of an asymptote at $\bar{\iota}$. The gradient is characteristic of logarithmic growth and not a constant asymptote. Thus, the assumption of the existence of an asymptote fails, and such a structure cannot exist. This is consistent with our physical intuition. \par
Therefore, we have shown that under considerations of no discrete changes in solutions, the formulation that begins by specifying the rotational transform is well-behaved. A numerical example of the possibility of specifying the problem in this or the standard way is shown in Figure \ref{fig:my_label}. A complete proof would place the smoothness assumption on more solid ground. \par
\begin{figure}
    \centering
    \includegraphics[width=0.45\textwidth]{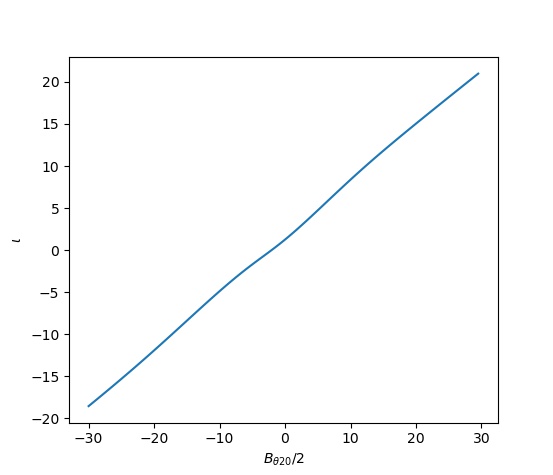}
    \caption{\textbf{Rotational transform as a function of the toroidal current $\bar{B}_{\theta20}$.} Example showing the relation between the rotational transform on axis and the toroidal current (roughly a measure of $\bar{Q}$) for solutions to the $\sigma$ Riccati equation. This shows how either formulation is a valid one numerically. The configuration corresponds to an axis defined by $R=1+0.22041\cos3\phi-0.012973\cos6\phi$ and $Z=0.22041\sin3\phi-0.012973\sin6\phi$, $\sigma(0)=0$, $\eta=\sqrt{2}$ and $B_0=1$. This corresponds to a QH configuration.}
    \label{fig:my_label}
\end{figure}
}
\par
\section{Magnetic shear}
In this Appendix, we present the closed forms of the various spatial functions needed for the calculation of the magnetic shear explicitly. As advertised in the main text, not all quantities through third order are required to compute the relevant quantities. We present the needed third-order quantities as well as the $\tilde{\Lambda}_3$ expression necessary for the evaluation of the magnetic shear. These are obtained by following the prescription detailed in the other Appendices, implemented into a symbolic handler (such as Mathematica). 
\begin{widetext}
\begin{align*}
    \tilde{\Lambda}_3=&\frac{2}{(Y_{11}^S)^2}\left[(B_{\alpha0}B_0+B_{\alpha0}B_{20})B_{\theta20}+\frac{1}{2}B_{\alpha0}B_{11}^CB_{\theta31}^C\right]+\\
    &+\frac{1}{Y^S_{1,1}{}^2} \left\{2 l'\left[\tau \left(X^C_{3,1}Y^S_{1,1}+2 X^C_{2,2}Y^S_{2,2}+X^C_{1,1}Y^S_{3,1}-X^S_{3,1}Y^C_{1,1}-2 X^S_{2,2}Y^C_{2,2}\right)+\kappa \left(2 X^C_{2,2}Z^S_{2,2}+X^C_{1,1}Z^S_{3,1}-2 X^S_{2,2}Z^C_{2,2}\right)\right]-\right.\\
    &\left.-2 \iota _0 \left(2 X^C_{2,2}{}^2+X^C_{1,1}X^C_{3,1}+2 X^S_{2,2}{}^2+2 Y^C_{2,2}{}^2+2 Y^S_{2,2}{}^2+Y^S_{1,1}Y^S_{3,1}+2 Z^C_{2,2}{}^2+2 Z^S_{2,2}{}^2\right)--X^S_{3,1}X^C_{1,1}{}'-2 X^S_{2,2}X^C_{2,2}{}'+\right.\\
    &\left.+2 X^C_{2,2}X^S_{2,2}{}'+X^C_{1,1}X^S_{3,1}{}'-Y^S_{3,1}Y^C_{1,1}{}'-2 Y^S_{2,2}Y^C_{2,2}{}'+2 Y^C_{2,2}Y^S_{2,2}{}'+Y^C_{1,1}Y^S_{3,1}{}'-2 Z^S_{2,2}Z^C_{2,2}{}'+2 Z^C_{2,2}Z^S_{2,2}{}'\right\}.
\end{align*}
\begin{align*}
    Z_{31}^C&=-\frac{1}{3 B_{\alpha0} X_{11}^C Y_{11}^S}\left[2 B_{\alpha0} X_{20} Y_{11}^C Z_{22}^S+4 B_{\alpha0} X_{22}^C Y_{11}^C Z_{22}^S-2 B_{\alpha0} X_{11}^C Y_{20} Z_{22}^S-4 B_{\alpha0} X_{11}^C Y_{22}^C Z_{22}^S+2 B_{\alpha0} X_{22}^C Y_{11}^S Z_{20}+\right.\\
    &\left.+2 B_{\alpha0} X_{11}^C Y_{22}^S Z_{20}-2 B_{\alpha0} X_{20} Y_{11}^S Z_{22}^C+4 B_{\alpha0} X_{11}^C Y_{22}^S Z_{22}^C-2 B_{\alpha0} X_{22}^S Y_{11}^C Z_{20}-4 B_{\alpha0} X_{22}^S Y_{11}^C Z_{22}^C-\right.\\
    &\left.-l(\phi ) \left(-2 \kappa B_{\psi0} X_{11}^C{}^2 Y_{11}^S+2 B_{\psi11}^C X_{11}^C Y_{11}^S+2 B_{\psi0} X_{11}^C Y_{22}^S+2 B_{\psi0} X_{22}^C Y_{11}^S-2 B_{\psi0} X_{22}^S Y_{11}^C+B_{\theta20} \left(2 X_{20} Y_{11}^C+X_{22}^C Y_{11}^C-\right.\right.\right.\\
    &\left.\left.\left.-2 X_{11}^C Y_{20}-X_{11}^C Y_{22}^C+X_{22}^S Y_{11}^S\right)+\tau \left(2 X_{20} Y_{11}^C+X_{22}^C Y_{11}^C-2 X_{11}^C Y_{20}-X_{11}^C Y_{22}^C+X_{22}^S Y_{11}^S\right)-2 \kappa X_{11}^C Z_{20}-\kappa X_{11}^C Z_{22}^C\right)+\right.\\
    &\left.+2 \iota _0 \left(X_{11}^C X_{22}^S-Y_{22}^C Y_{11}^S+Y_{11}^C Y_{22}^S\right)+2 X_{11}^C  X_{20}'+X_{11}^C X_{22}^C{}'+2 Y_{11}^C Y_{20}'+Y_{11}^C Y_{22}^C{}'+Y_{11}^S Y_{22}^S{}'\right]
\end{align*}
\begin{align*}
    Z_{31}^S&=\frac{1}{3 B_{\alpha0} X_{11}^C Y_{11}^S}\left[-2 B_{\alpha0} X_{22}^C Y_{11}^C Z_{20}+2 B_{\alpha0} X_{11}^C Y_{22}^C Z_{20}+2 B_{\alpha0} X_{20} Y_{11}^C Z_{22}^C-2 B_{\alpha0} X_{11}^C Y_{20} Z_{22}^C+2 B_{\alpha0} X_{20} Y_{11}^S Z_{22}^S-\right.\\
    &\left.-4 B_{\alpha0} X_{22}^C Y_{11}^S Z_{22}^S-2 B_{\alpha0} X_{22}^S Y_{11}^S Z_{20}+4 B_{\alpha0} X_{22}^S Y_{11}^S Z_{22}^C+l(\phi ) \left(2 B_{\psi0} \left(X_{22}^C Y_{11}^C-X_{11}^C Y_{22}^C+X_{22}^S Y_{11}^S\right)+2 B_{\psi11}^S X_{11}^C Y_{11}^S+\right.\right.\\
    &\left.\left.+2 B_{\theta20} X_{20} Y_{11}^S-B_{\theta20} X_{22}^C Y_{11}^S-B_{\theta20} X_{11}^C Y_{22}^S+B_{\theta20} X_{22}^S Y_{11}^C+\tau \left(2 X_{20} Y_{11}^S-X_{22}^C Y_{11}^S-X_{11}^C Y_{22}^S+X_{22}^S Y_{11}^C\right)-\kappa X_{11}^C Z_{22}^S\right)-\right.\\
    &\left.X_{11}^C X_{22}^S{}'+2 \iota_0 \left(X_{11}^C X_{22}^C+Y_{11}^C Y_{22}^C+Y_{11}^S Y_{22}^S\right)-2 Y_{11}^S Y_{20}'+Y_{11}^S Y_{22}^C{}'-Y_{11}^C Y_{22}^S{}'\right]
\end{align*}
\begin{align*}
    X_{31}^C&=\frac{1}{2 l^2 \kappa}\left[B_{\alpha0}^2 \left(-B_{31}^C\right)-2 B_{\alpha 1} B_{\alpha0} B_{11}^C+l^2 \kappa^2 X_{11}^C \left(2 X_{20}+X_{22}^C\right)+2 l^2 \tau^2 X_{11}^C X_{20}+l^2 \tau^2 X_{11}^C X_{22}^C+2 l^2 \tau^2 Y_{11}^C Y_{20}+\right.\\
    &\left.+l^2 \tau^2 Y_{11}^C Y_{22}^C+l^2 \tau^2 Y_{11}^S Y_{22}^S+2 l \tau Y_{20} X_{11}^C{}'+l \tau Y_{22}^C X_{11}^C{}'+2 l \tau Y_{11}^C X_{20}'+l \tau Y_{11}^C X_{22}^C{}'-2 l \tau X_{20} Y_{11}^C{}'-l \tau X_{22}^C Y_{11}^C{}'-2 l \tau X_{11}^C Y_{20}'-\right.\\
    &\left.-l \tau X_{11}^C Y_{22}^C{}'+l \kappa \left(l \tau \left(Y_{11}^C \left(2 Z_{20}+Z_{22}^C\right)+Y_{11}^S Z_{22}^S\right)+2 Z_{20} X_{11}^C{}'+Z_{22}^C X_{11}^C{}'-2 X_{11}^C Z_{20}'-X_{11}^C Z_{22}^C{}'-3 \iota _0 X_{11}^C Z_{22}^S\right)-\right.\\
    &\left.-2 \iota _0 l \tau X_{20} Y_{11}^S-3 \iota _0 l \tau X_{22}^C Y_{11}^S-3 \iota _0 l \tau X_{11}^C Y_{22}^S+3 \iota _0 l \tau X_{22}^S Y_{11}^C+l \tau Y_{11}^S X_{22}^S{}'-l \tau X_{22}^S Y_{11}^S{}'+2 l Z_{31}^C{}'+2 \iota _0 l Z_{31}^S+2 \iota _0^2 X_{11}^C X_{22}^C+\right.\\
    &\left.-2 \iota _0 X_{22}^S X_{11}^C{}'-\iota _0 X_{11}^C X_{22}^S{}'+2 X_{11}^C{}' X_{20}'+X_{11}^C{}' X_{22}^C{}'+2 \iota _0^2 Y_{11}^C Y_{22}^C+2 \iota _0 Y_{22}^S Y_{11}^C{}'+2 \iota _0 Y_{11}^S Y_{20}'+\iota _0 Y_{11}^S Y_{22}^C{}'-2 \iota _0 Y_{22}^C Y_{11}^S{}'-\right.\\
    &\left.-\iota _0 Y_{11}^C Y_{22}^S{}'+2 Y_{11}^C{}' Y_{20}'+Y_{11}^C{}' Y_{22}^C{}'+2 \iota _0^2 Y_{11}^S Y_{22}^S+Y_{11}^S{}' Y_{22}^S{}'\right]
\end{align*}
\begin{align*}
    X_{31}^S&=\frac{1}{2 l^2 \kappa}\left[-B_{\alpha0}^2 B_{31}^S+l^2 \kappa^2 X_{11}^C X_{22}^S+l^2 \tau^2 X_{11}^C X_{22}^S+2 l^2 \tau^2 Y_{20} Y_{11}^S-l^2 \tau^2 Y_{22}^C Y_{11}^S+l^2 \tau^2 Y_{11}^C Y_{22}^S+l^2 \kappa \tau Y_{11}^C Z_{22}^S+\right.\\
    &\left.=2 l^2 \kappa \tau Y_{11}^S Z_{20}-l^2 \kappa \tau Y_{11}^S Z_{22}^C+\iota _0 \left(l \tau \left(2 X_{20} Y_{11}^C-3 X_{22}^C Y_{11}^C-2 X_{11}^C Y_{20}+3 X_{11}^C Y_{22}^C-3 X_{22}^S Y_{11}^S\right)+l \kappa X_{11}^C \left(3 Z_{22}^C-2 Z_{20}\right)-\right.\right.\\
    &\left.\left.-2 l Z_{31}^C-2 X_{22}^C X_{11}^C{}'-2 X_{11}^C X_{20}'+X_{11}^C X_{22}^C{}'-2 Y_{22}^C Y_{11}^C{}'-2 Y_{11}^C Y_{20}'+Y_{11}^C Y_{22}^C{}'-2 Y_{22}^S Y_{11}^S{}'+Y_{11}^S Y_{22}^S{}'\right)+l \tau Y_{22}^S X_{11}^C{}'+\right.\\
    &\left.+2 l \tau Y_{11}^S X_{20}'-l \tau Y_{11}^S X_{22}^C{}'-2 l \tau X_{20} Y_{11}^S{}'+l \tau X_{22}^C Y_{11}^S{}'-l \tau X_{11}^C Y_{22}^S{}'+l \kappa Z_{22}^S X_{11}^C{}'-l \kappa X_{11}^C \text{Zs}_{2,2}'(\phi )+l \tau Y_{11}^C X_{22}^S{}'-\right.\\
    &\left.-l \tau X_{22}^S Y_{11}^C{}'+2 l \text{Zs}_{3,1}'(\phi )+2 \iota _0^2 \left(X_{11}^C X_{22}^S-Y_{22}^C Y_{11}^S+Y_{11}^C Y_{22}^S\right)+X_{11}^C{}' X_{22}^S{}'+2 Y_{20}' Y_{11}^S{}'-Y_{22}^C{}' Y_{11}^S{}'+Y_{11}^C{}' Y_{22}^S{}'\right]
\end{align*}
\begin{align*}
    Y_{31}^S&=\frac{1}{4 B_{\alpha0} X_{11}^C}\left[4 B_{\alpha0} \left(-X_{31}^C Y_{11}^S-2 X_{22}^C Y_{22}^S+X_{31}^S Y_{11}^C+2 X_{22}^S Y_{22}^C\right)-2 B_{\alpha 1} X_{11}^C Y_{11}^S+2 B_{\psi0} Y_{11}^S X_{11}^C{}'+2 B_{\psi0} X_{11}^C Y_{11}^S{}'-\right.\\
    &\left.-l(\phi ) \left(\tau B_{\theta20} \left(X_{11}^C{}^2+Y_{11}^C{}^2+Y_{11}^S{}^2\right)+4 \kappa X_{20}\right)-B_{\theta20} Y_{11}^C X_{11}^C{}'+B_{\theta20} X_{11}^C Y_{11}^C{}'+2 \iota _0 B_{\theta20} X_{11}^C Y_{11}^S+4 Z_{20}'\right]
\end{align*}
\end{widetext}
We have taken i) from $C_\perp^2$ the functions $Z_{31}^C$ and $Z_{31}^S$, ii) from $J^3$ functions $X_{31}^C$ and $X_{31}^S$ (which will explicitly introduce the third-order magnetic corrections $B_{31}^C$ and $B_{31}^S$), and iii) from $C_b^2$ the function $Y_{31}^S$. These expressions are involved by the very nature of the many orders gathered in the expansion.

\section*{Acknowledgements}
We are grateful to J. Burby, N. Duignan, and J. Meiss for fruitful discussions. This research is primarily supported by a grant from the Simons Foundation/SFARI (560651, AB) and DoE Contract No DE-AC02-09CH11466. ER was also supported by the Charlotte Elizabeth Procter Fellowship at Princeton University.

\section*{Data availability}
The data that support the findings of this study are available from the corresponding author upon reasonable request.

\bibliography{weakQSnaeProof}% Produces the bibliography via BibTeX.

\end{document}